\documentclass[12pt,a4paper]{article}

\usepackage{graphicx}
\usepackage{setspace} 
\usepackage{amsmath} 
\usepackage{titling} 

\usepackage{color}

\usepackage{hyperref}
\hypersetup{unicode=true,
           colorlinks=true,
	   linkcolor=blue,
	   citecolor=red,
           pdfborder={0 0 0}
}

\newcommand{\tB}{\ensuremath t_\mathrm{B}}
\newcommand{\Glin}{\ensuremath G_\mathrm{lin}}
\newcommand{\Vg}{\ensuremath V_\mathrm{g}}

\newcommand{\nm}{\ensuremath\mathrm{nm}}

\newcommand{\eV}{\ensuremath \mathrm{eV}}

\begin{document}
\title{Conductance of graphene flakes contacted at their corners}
\author{%
	Martin Kon\^{o}pka
	\\ \\
	Department of Physics,\\
	Institute of Nuclear and Physical Engineering,\\
	Faculty of Electrical Engineering and Information Technology,\\
	Slovak University of Technology in Bratislava,\\
	Ilkovi\v{c}ova 3, 812~19 Bratislava, Slovakia
	\\
	\small{E-mail: \href{mailto:martin.konopka@stuba.sk}{martin.konopka@stuba.sk}}}
\date{7$^\textrm{th}$ October 2015}
\maketitle
\begin{abstract}
\noindent
Linear conductance of junctions formed by graphene flakes with order of nanometer-thick electrodes attached
at the corners of the flakes is studied.
The explored structures have sizes up to {20\,000} atoms and the conductance is studied
as a function of applied gate voltage varied around the Fermi level.
The finding, obtained computationally, is that junctions formed by armchair-edge flakes with the electrodes connected
at the acute-angle corners block the electron transport while only junctions with such electrodes
at the obtuse-angle corners tend to provide the high electrical conductance typical for metallic GNRs.
The finding in case of zig-zag edges is similar with an exception of a relatively narrow gate voltage interval
in which each studied junction is highly conductive as mediated by the edge states.
The contrast between the conductive and insulating setups is typically several orders of magnitude
in terms of ratio of their conductances.
Main results of the paper remain to a large extent valid also in the presence of edge disorder.
\end{abstract}
\begin{small}
\quote{\textbf{Keywords:} graphene, conductance, transport, edge, corner, gate voltage, flakes}
\begin{quotation}
\noindent
This is an author-created, un-copyedited version of an article published in \textit{Journal of Physics: Condensed Matter}.
IOP Publishing Ltd is not responsible for any errors or omissions in this version of the manuscript or any version derived from it.
The Version of Record is available online at:
\begin{center}
\href{http://dx.doi.org/10.1088/0953-8984/27/43/435005}{\textcolor{blue}{http://dx.doi.org/10.1088/0953-8984/27/43/435005}}
\end{center}
\end{quotation}
\end{small}
%
%
\section{\label{sec:intro} Introduction}
%
%
Extensive research of graphene was initiated by its unique electronic properties and at the same time by the 
possibility to prepare graphene samples in laboratories~\cite{Wallace,N_G_review}.
On the theoretical side, the dc electronic transport in ideal graphene and GNRs is the process of fundamental interest.
Experiments addressing electronic transport in graphene necessarily employ finite graphene flakes, usually rectangular
graphene nano-ribbons (GNRs).
Therefore,
in GNRs, their finite spatial dimensions and consequently the boundary conditions must be taken into account.
It is known that armchair GNRs (AGNRs) can either be metallic or semi-conductive,
depending on their width~\cite{Nakada96,Louie_PRL06}.
The zig-zag GNRs (ZGNRs) are always metallic and support special localised states at their edges~\cite{Nakada96}.
Contacts of the electrodes to a GNR are often assumed to extend across the entire width of the ribbon.
While the conductance properties of such junctions
have been frequently studied~\cite{N_G_expt,Katsnelson06,Beenakker06}
(see also review~\cite{N_G_review} and references therein),
little is known about graphene flakes contacted at their corners.
In their computational work~\cite{Cuong_2013} Cuong \textit{et al.}
reported the absence of edge states near the $120^{\circ}$ corners of ZGNRs.
This finding is confirmed also by our calculations and will be referenced below in the text.
In another computational study Borunda \textit{et al.} investigated current flow through rectangular
GNRs assuming electrodes contacted at certain positions at opposite edges of the GNR~\cite{Borunda13}.
Based on their simulations the authors were able to differentiate ballistic and diffusive regimes of transport
in graphene.
Related to our present work is also the theoretical analysis of a dual-probe STM setup on graphene
by Settnes \textit{et al.}~\cite{Settnes14}.
The authors computed conductance maps assuming the two STM tips at various positions on graphene surface, including
non-ideal samples.
We however could not find any analysis of graphene junctions with order of nanometer thick electrodes attached at the
corners of the flake.
Therefore the goal of this work is to present our results on the dc linear conductance of graphene flakes of various
shapes and edges and with the electrodes connected at the corners of the flakes.
Particular investigated samples have the shapes of trapezoids, triangles, rhombi and rhomboids,
all with angles at corners being integer multiplies of $30^\circ$.
We find that the conductance strongly depends on the angles at the corners
to which the electrodes are connected.
Acute-angle corners ($30^\circ$ and $60^\circ$ studied here) in most cases do not enable good electronic transport.
The opposite is true for electrodes attached at obtuse-angle corners ($120^\circ$ studied here):
they often provide high conductance levels typical for metallic GNRs.
We describe GNRs' electronic structure using a tight-binding (TB) approximation with hoppings
up to third nearest neighbor (NN) as specified below in section~\ref{sec:methods}.
The model of the electrodes uses the first NN approximation and will further be described as well.
Main results of the paper are explained within section~\ref{sec:results}, considering graphene flakes with shapes
of symmetrical trapezoids and triangles.
Section~\ref{sec:LDOS} supports the main results by a local density of states analysis.
Section~\ref{sec:concl} provides a summary and discussion of the results.
Comparisons to results obtained within the first NN TB approximation (1NNTBA),
further details and examination of the model of the electrodes,
and a section describing results for samples with edge disorder
are supplied as appendices.
Supplementary information (SI) contains results for graphene flakes of several additional shapes including
a system with wide contacts to electrodes, further results within the 1NNTBA and a section with data for the flakes
terminated by imperfect edges.
%
%
\section{\label{sec:methods} Models and Methods}
%
%
Our model of GNR's electronic structure employs one explicit electron per atom,
the independent-electron approximation and a TB Hamiltonian with hoppings
up to the third NN included (3NNTBA).
%
\begin{figure}[!t]
\centerline{\includegraphics[width=0.49\textwidth]{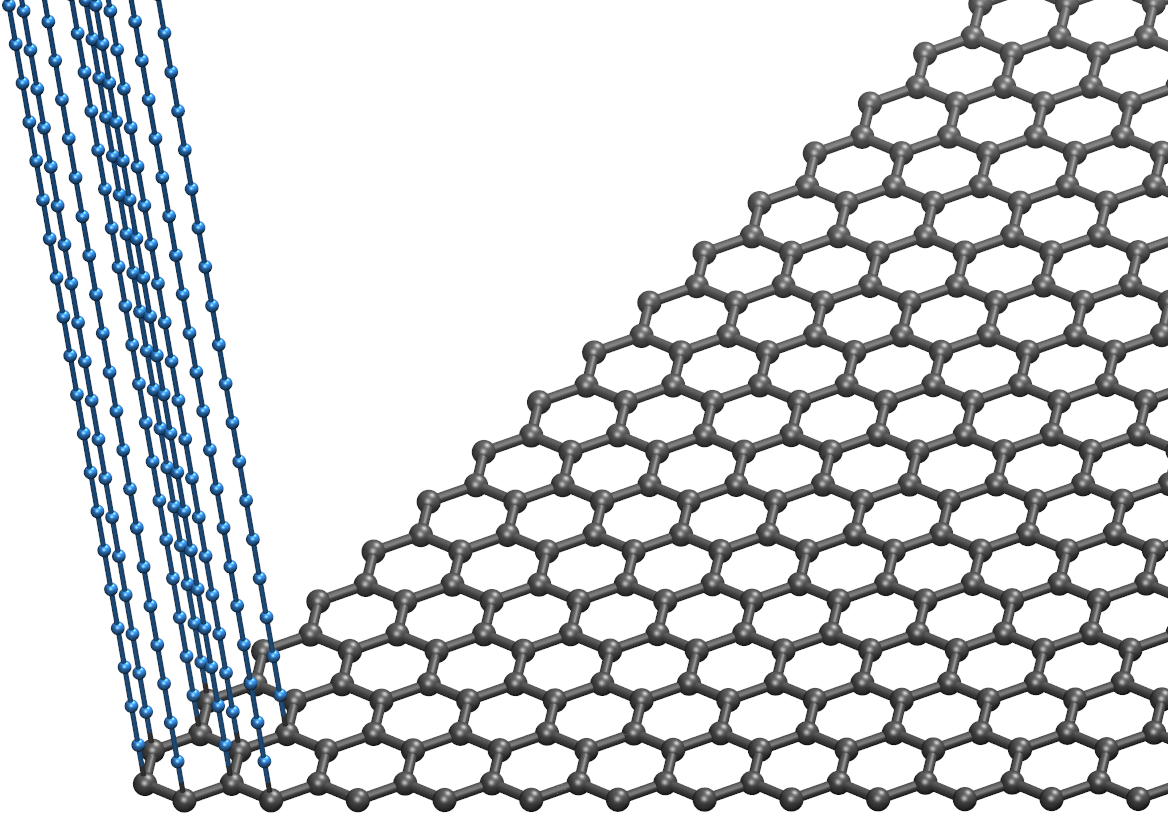}}
\caption{Schematic graphical representation of our model of the electrodes and their coupling to the graphene flake.
In this example the flake has zig-zag edges and the electrode is attached to its $60^\circ$ corner.
The electrode itself is formed by a bunch of monoatomically thin semi-infinite mutually decoupled wires.
In typical setups we use 35-36 wires per electrode.
Here we show only 12 of them for easier visualisation.
The interatomic distances within each wire are $a$, i.e. the same as is the NN distance in graphene.
See also appendix~\ref{sec:electrodes} in which we consider a model with wires mutually coupled along
a finite length.}
\label{fig:indep_wires}
\end{figure}
%
According to~\cite{Reich2002}, we set the hopping parameters to values
$\tB = -2.97\,\eV$, $\tB' = -0.073\,\eV$ and $\tB'' = -0.33\,\eV$.
Magnitude of $\tB$ is conveniently used as the energy unit.
Interatomic overlaps are neglected.
The edges of the flakes are assumed to be terminated by hydrogen atoms electrons of which do not explicitly
enter the TB Hamiltonian; see for instance~\cite{White07}.
Although the TB modelling is a relatively basic-level one, it has, even in the 1NNTBA,
in many cases been proven to provide satisfactory description of conductance properties of graphene structures
on at least semi-quantitative level which is sufficient for the present study.
Such models are frequently employed in theoretical studies of graphene samples~\cite{Beenakker06,Malysheva08,Perfetto10}
and carbon nanotubes~\cite{Ando05}.

In the present study the whole system including the electrodes is composed of identical atoms.
Each of the two electrodes in our model is composed of a number (from tens to hundreds)
of monoatomically thin mutually \emph{non-interacting} wires (figure~\ref{fig:indep_wires}).
Within each wire the 1NNTBA with the parameter $\tB$ is used to describe the electronic structure.
This simple model of the electrodes was found very convenient in~\cite{our_PRB_2014}
where it was used for GNRs and provided results in a semi-quantitative agreement with different models
of the electrodes~\cite{Rosenstein09,Perfetto10}.
The coupling of the wires to the GNR is again described by the first NN model using the parameter $\tB$.
See appendix~\ref{sec:electrodes} where we provide an examination and justification of such model of the electrodes
for the use in the present work.

Each atom of the system is characterised by its on-site energy.
In equilibrium and in absence of any external field these energies are set to zeros for all the atoms of the entire
system.
The effect of the applied bias voltage $U$ as well as of a gate voltage $\Vg$ are modelled by variations
of the on-site energies of the atoms in the electrodes and/or in the graphene flake~\cite{our_PRB_2014}.
Variation of the gate voltage represents the shift of the chemical potential in the graphene flake
away from the neutrality point and is considered within a $0.2\,|\tB|$ wide interval around the Fermi level.
We apply almost limitingly small (still numerically finite) values of the bias voltages $U$, compute resulting
dc currents $I$ through the junctions and consequently obtain the linear conductances $\Glin = \lim_{U \to 0} I/U$.
The considered model can now be briefly summarised:
Hamiltonian of the entire system is
\begin{equation}
\hat{H} = \sum_{l,l'} H_{l,l'} \, a^\dag_l a_{l'}
\, ,
\label{eq:ham}
\end{equation}
with $l$ and $l'$ running over atomic sites of the entire system (including the electrodes) and
$a^\dag_l$, $a_{l'}$ being the creation and annihilation operators of an electron in the TB orbitals
$|l\rangle$ and $|l'\rangle$, respectively.
It is assumed that the TB orbitals form and orthonormal basis set: $\langle l|l'\rangle = \delta_{ll'}$.
The matrix elements of the Hamiltonian in our model are set to
\begin{equation}
H_{l,l'} = \left\{
\begin{array}{ll}
e U			& l = l' \ \textrm{within the wires of the left electrode}\\
0			& l = l' \ \textrm{within the wires of the right electrode}\\
(e \Vg + e U/2)		& l = l' \ \textrm{within the flake}\\
\tB			& l, l' \ \textrm{being any first NN}\\
\tB'			& l, l' \ \textrm{being second NN within the flake}\\
\tB''			& l, l' \ \textrm{being third NN within the flake}\\
0			& \textrm{all other} \ l, l'
\end{array}
\right.
\ .
\label{eq:matel}
\end{equation}
Since the monoatomically thin wires are assumed to be mutually decoupled, matrix elements $H_{l,l'}$ with
$l$, $l'$ belonging to different wires are all zeros, even if the two wires belong to same electrode.
The exceptions are the examples used in appendix~\ref{sec:electrodes}
in which we examine a more complex model of the electrodes.
As mentioned above, we explore the linear regime, choosing particular numerical value of the bias voltage
$U = 10^{-9}\,|\tB|/e$.
Around and below this magnitude of $U$ the $I(U)$ function is perfectly linear;
even higher values would still provide the linear regime.
The contribution $e U/2$, although applied within the graphene flake [line 3 in equation~\eqref{eq:matel}],
does not have a real effect in this particular study because of the small value of $U$.
We note in passing that application of the 1NNTBA in the entire system would yield
a perfectly symmetric functional dependence of $\Glin$:  $\Glin(-\Vg) = \Glin(\Vg)$, thanks to the symmetry
of the dispersion relations of the wires as well as of graphene~\cite{Wallace}.
Inclusion of the hopping matrix elements up to the third NN in graphene flakes breaks the symmetry.

In addition,
the extension of the TB hopping range in the graphene flakes causes, among other features, a horizontal shift of the
calculated $\Glin(\Vg)$ profile compared to the results from the 1NNTBA.
This shift results from the approximate model~\eqref{eq:matel}.
Its value would change with inclusion of overlaps in the TB model.
To facilitate presentation of the results we apply a proper compensating horizontal shift
$\Delta\Vg = +0.072\,|\tB|/e$
to all plots obtained by the 3NNTBA
so that they match the symmetry of corresponding 1NNTBA curves as much as possible
in cases of the armchair-edged structures.
We apply the same compensating shift $\Delta\Vg$ also for zig-zag-edged flakes
although in this case the plots can not be made even approximately symmetric.
Hence our horizontal axes description in terms of quantities formally equals to $\Vg + \Delta\Vg$.

Electron currents in the present work are obtained using a Landauer-type formula in the same way as we used
in stationary calculations in~\cite{our_PRB_2014}.
Because an electrode in our work is modelled as a bunch of mutually non-interacting monoatomic wires,
we formally have a multi-terminal system and use a multi-terminal formalism.
For example, each of the junctions depicted in figure~\ref{fig:trap_ac} uses two 36-wires thick electrodes.
Such junctions are formally treated as 72-terminal systems within our approach.
We provide more details on the model of the electrodes in appendix~\ref{sec:electrodes} where we also demonstrate
that the model based on the non-interacting wires does not yield any significant consequences to the
calculated linear conductances.
The core part of the multi-terminal formalism are the quantum-mechanical calculations which employ
the scattering approach with wavefunctions obtained from
Green function formalism~\cite{Ryndyk}, again in the same way as we used in~\cite{our_PRB_2014}.
A scattering approach and the Landauer-type formula were used also in our work~\cite{our_EPJB} for systems with just
two terminals; see equations~(24) and (25) therein.\footnote{Formula~(24) in our work~\cite{our_EPJB} has the
density-of-states factors included in the definitions of coefficients $\mathcal{A}$ and $\mathcal{C}$.}
Here we use the multi-terminal generalisation of the method and instead of the simple scattering approach
we employ the more general one based on the Green function technique which allows us to obtain the wavefunctions.
%
%
\section{\label{sec:results} Results}
%
%
Conductance properties of a graphene-flake formed nano-junction depend on a number of parameters and conditions:
on the intrinsic properties of the flake itself (its shape, size)
and also on the way how and where the electrodes are attached to the flake.
Among the plethora of possible scenarios we focus on several representative ones with the aim to study junctions
with the electrodes at the corners.
We present our results in figures which display particular considered structures together with the linear conductance
$\Glin$ as a function of the chemical potential variation (here represented by the gate voltage)
around the neutrality point.
Results for a limited set of structures only are described in the text,
which however should be sufficient to explain our main findings.
Results for other structures can be found in the SI.
In our description of the physical structure sizes we often employ the graphene NN distance $a \approx 1.42\,\nm$
as well as its lattice parameter $b = a \sqrt{3} \approx 2.46\,\nm$.
%
\subsection{Armchair edges}
\begin{figure}[!t]
\centerline{\includegraphics[width=0.90\textwidth]{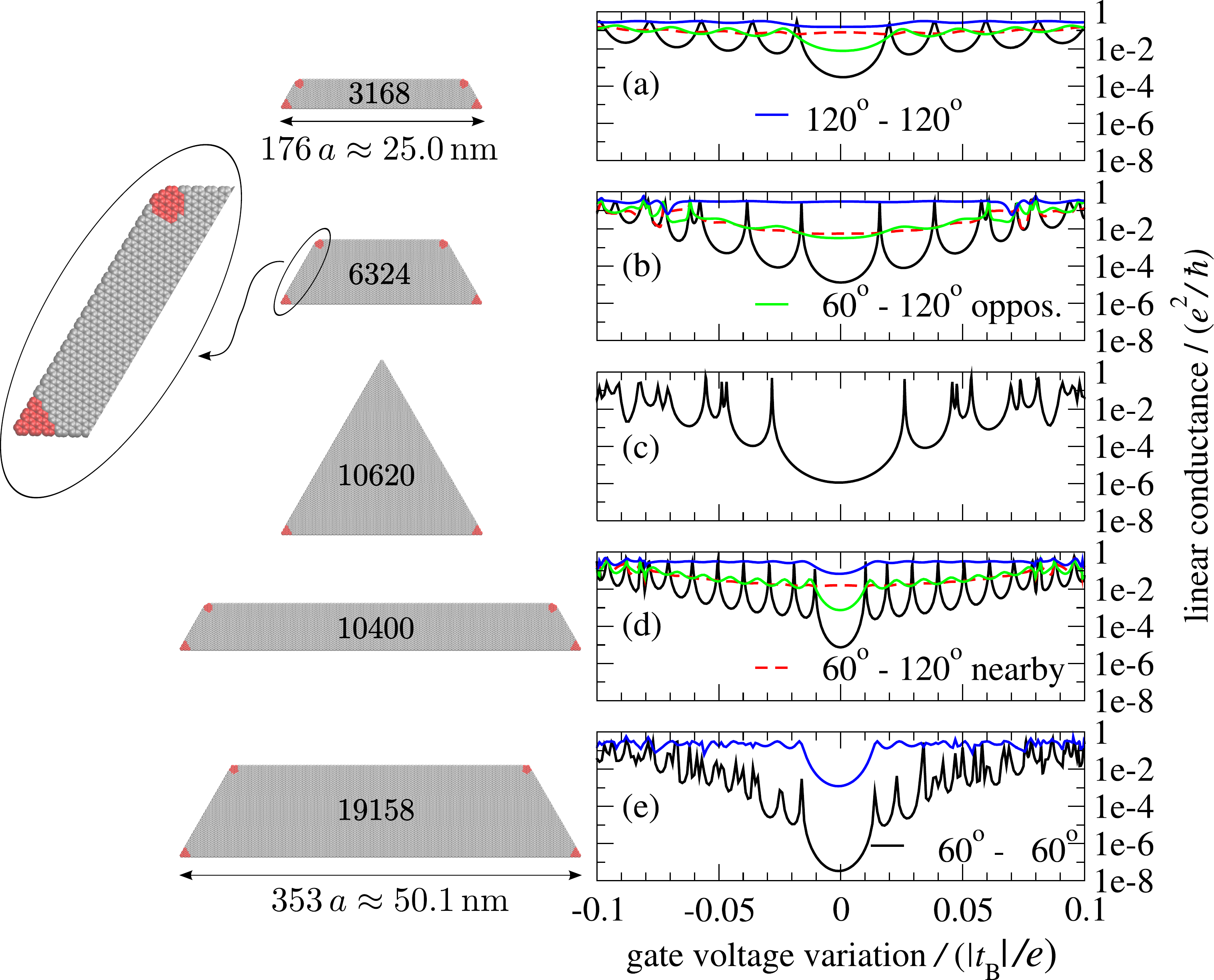}}
\caption{Results for the trapezoidal flakes with armchair edges.
\textbf{Left panel:}
graphical representation of the flakes.
Red-colored atoms in the corners are directly coupled to the electrodes.
Each individual considered setup employs just two of the four colored corners serving as the electrodes-attachment
areas.
The electrodes are $36$-monoatomic wires thick.
The numbers of atoms in each flake are displayed by labels $3168$, $\dots$, $19\,158$.
$a \approx 1.42\,\nm$ is the NN distance in graphene.
See main text for more details.
\textbf{Right panel:}
linear conductances $\Glin$ plotted as functions of the gate voltage for the structures shown in the left panel
in the same vertical order.
$\tB$ is the nearest-neighbor tight-binding parameter used for the entire system and $e$ is the unit charge.
Solid black curves: both electrodes at the $60^\circ$ corners.
Dashed red curves: one electrode at the $60^\circ$ corner, the other at the nearby $120^\circ$ corner.
Solid green curves: one electrode at the $60^\circ$ corner, the other at the opposite $120^\circ$ corner.
Solid blue curves: both electrodes at the $120^\circ$ corners.
Legends like $120^\circ-120^\circ$ apply globally within the whole figure.
For graph (e) the two plots only have been calculated.
The independent variable on the horizontal axes is $\Vg + \Delta\Vg$, with $\Delta\Vg = +0.072\,|\tB|/e$ as
explained in section~\ref{sec:methods}.}
\label{fig:trap_ac}
\end{figure}

%
As representative structures with armchair (ac) edges
we consider symmetrical trapezoids with acute angles of $60^\circ$ at their base
(left panel of figure~\ref{fig:trap_ac}).
%
%
These flakes support four different basic attachments of the electrode pair:
(i)~both electrodes at the acute angles (AA-AA setups),
(ii)~one electrode at the acute angle and the other at the nearby obtuse angle (AA-OA setup),
(iii)~one electrode at the acute angle and the other at the opposite obtuse angle (the other AA-OA setup),
(iv)~both electrodes at the obtuse angles (OA-OA setup).
The top three structures in figure~\ref{fig:trap_ac} share the same base length $L = 176\,a \approx 25.0\,\nm$.
The equilateral triangle in this context can be considered as a limiting-case trapezoid.
The two trapezoids at the bottom represent wider and overall larger structures ($L = 353\,a \approx 50.1\,\nm$)
and allow us to observe size-dependence of the computed results.
The heights of the samples are $14\,b$, $32\,b$, $88\,b$, $23\,b$ and $46\,b$
from the upper-most structure to the one at the bottom.
Total numbers of atoms composing given flake are typed on the images.
Each electrode considered for figure~\ref{fig:trap_ac} is 36 monoatomic wires thick.
While possible, we choose the shape of the contact areas the same for corresponding
corners.\footnote{The largest trapezoid has sharper corners at its obtuse angles, compared to the smaller structures.}
All trapezoids in figure~\ref{fig:trap_ac} have (vertical) widths corresponding to metallic
AGNRs~\cite{Nakada96,Louie_PRL06}.
Right panel of the figure displays our results expressed in terms of the linear conductance
as a function of the gate voltage $\Vg + \Delta\Vg$.
Importantly,
we find most of the setups with both electrodes at the $60^\circ$ corners (the AA-AA junctions)
to be relatively insulating near the neutrality point
(black solid curves in figure~\ref{fig:trap_ac})
with their linear conductances $\Glin$ being several orders of magnitude below those for OA-OA setups
(blue solid curves).
The current-blocking behaviour is mostly retained also at elevated chemical potentials apart from the isolated peaks.
On the contrary,
configurations $120^\circ - 120^\circ$ (the blue plots) are highly conductive,
with $\Glin \approx 0.2-0.3\,e^2/\hbar$ roughly independent on the gate voltage,
with the exception of the largest flake [figure~\ref{fig:trap_ac}(e)];
even the latter example provides a huge contrast (4-5 orders in magnitude) between the conductances of the OA-OA and
AA-AA setups at the Fermi level.
The order of $10^{-1}\,e^2/\hbar$ conductance magnitudes are typical for metallic graphene
nano-ribbons~\cite{Katsnelson06,Beenakker06}.
The case of the mixed setups (one electrode at the $60^\circ$ angle while the other at one of the $120^\circ$ angles)
is more complicated;
see dashed red and solid green plots in figure~\ref{fig:trap_ac}.
Typically, linear conductances of these AA-OA setups take intermediate values.
An interesting feature in some of the AA-OA setups, found for example in figure~\ref{fig:trap_ac}(b),
is that the profile of the $\Glin$ function is mostly independent
on the choice of the particular pair of the AA-OA attachment corners:
both the nearby-corners case and the opposite-corners case yield similar curves;
compare the dashed red and solid green plots on the figures.
The similarity is reduced with increasing relative length of trapezoids.
This is an intuitively comprehensible feature because for long-narrow trapezoids
the mutual positions of the contacts in the two $60^\circ-120^\circ$ setups are quite different.
In SI we show that there is no such similarity in case of ac-edge terminated long-narrow rhomboids.
Note that the AA-OA junctions have not been calculated for the largest trapezoid, figure~\ref{fig:trap_ac}(e).
Finally we refer to appendix section~\ref{sec:1st3rdNN} which provides results employing the linear scale on the
conductance axis~[figures~\ref{fig:1st_3rd_NN_ac_overall} and~\ref{fig:1st_3rd_NN_ac_detail}].
%
\subsection{\label{ssec:ZZE} Zig-zag edges}
%
For graphene flakes with ZZ edges
%
\begin{figure}[!t]
\centerline{\includegraphics[width=0.773\textwidth]{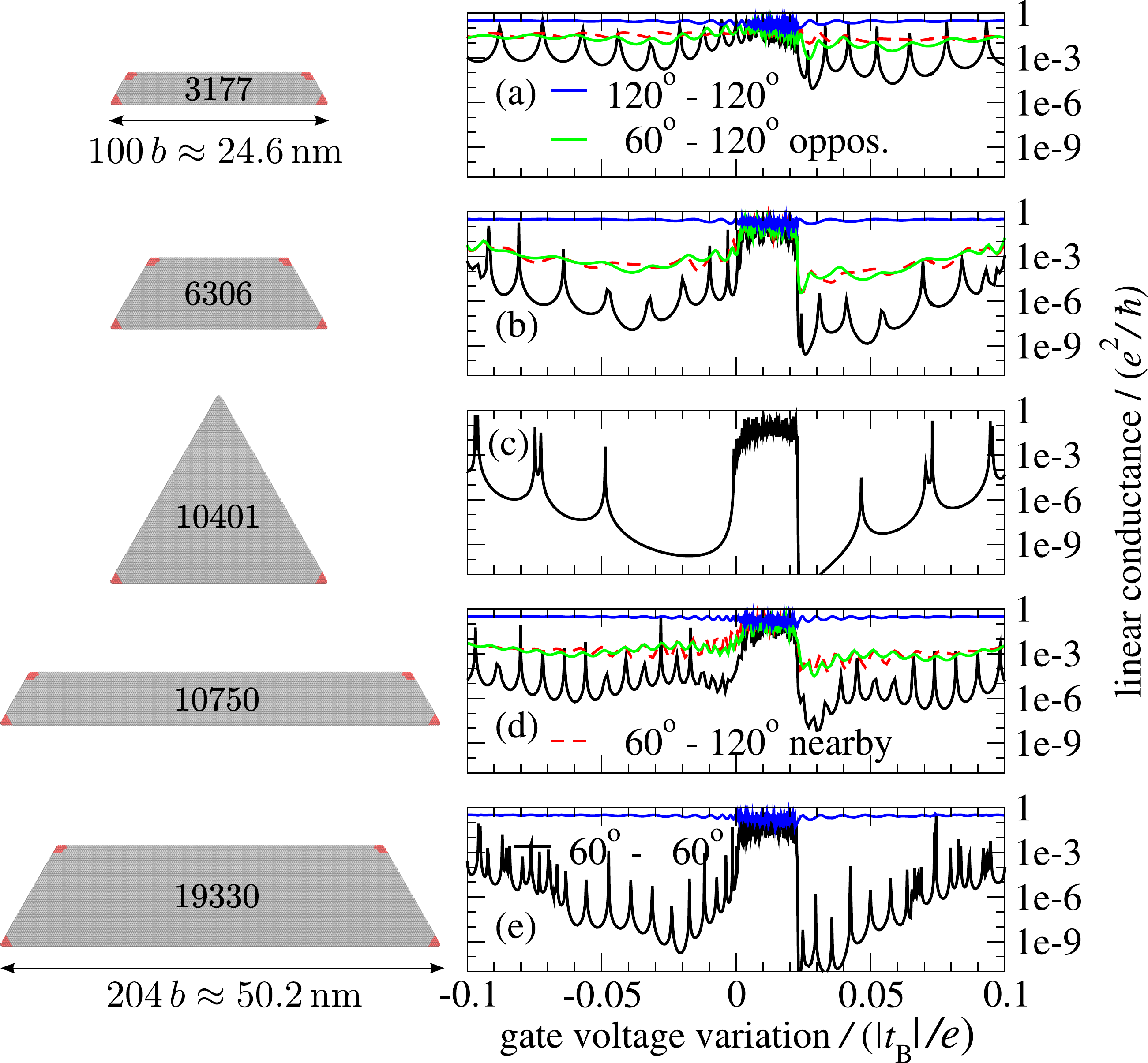}}
\caption{Results for the flakes with zig-zag edges.
All description as well as the color coding is analogous to that in figure~\ref{fig:trap_ac}.
The electrodes are 35-monoatomic wires thick now.
Here as well as in all other graphs displaying the conductances obtained within the 3NNTBA model
we apply a uniform horizontal shift
$\Delta\Vg = +0.072\,|\tB|/e$ to all computed plots as described in section~\ref{sec:methods}.
The distance $b$ is the lattice parameter of graphene given by the relation $b = a \sqrt{3} \approx 2.46\,\nm$,
where $a \approx 1.42\,\nm$ is the NN distance in graphene.}
\label{fig:trap_zz}
\end{figure}
%
we perform an analysis analogous to that in previous section, choosing now electrodes $35$ monoatomic wires thick.
We again choose structures of trapezoidal shapes.
The samples displayed in figure~\ref{fig:trap_zz} are of two different lengths, $100\,b$ and $204\,b$.
Their heights from the upper-most flake to the lowest one are $24.5\,a$, $56\,a$, $150.5\,a$, $41\,a$ and $80\,a$.
Generally valid differences if compared to the ac-terminated flakes (figure~\ref{fig:trap_ac}) are that
(i)~The AA-AA junctions now provide
a significant conductance within a $0.025\,|\tB|$ narrow gate voltage window;
this is an effect of the special localised ZZ-edge state and can be found significant
if TB hoppings up to the third NN are included;
this high-conductance regime would not be found for AA-AA setups within just the 1NNTBA
as we have checked and provide several comparisons in appendix~\ref{sec:1st3rdNN}, most relevantly
in figure~\ref{fig:1st_3rd_NN_zz_detail}(b).
(ii)~The conductance ratios of the OA-OA setups and the remaining ones are now larger for most of the studied $\Vg$
range.
(iii)~The $\Glin$ spectra are significantly asymmetric around the Fermi level.
(iv)~Rapid oscillations are visible in the $\Glin(\Vg)$ spectra within the high conductance window.

As mentioned above, the high conductance regime
of the AA-AA setup
found in the vicinity of the Fermi level
comes from the contribution of the ZZ edge state.
figure~\ref{fig:trap_zz} shows the rapid oscillations of $\Glin$ in this gate voltage range.
The oscillations can be more clearly seen in data obtained within the 1NNTBA and consequently
more easily interpreted.
We demonstrate such results in SI for long-narrow rhomboids.
The presence of the oscillations even at gate voltages arbitrarily close to the zero
can be understood taking into account the effect of the electrodes and coupling of their modes to the flake:
the semi-infinite electrodes support modes at energies arbitrarily close to the Fermi level.

Away from the energies or in absence of the ZZ-edge induced peaks,
the electronic transport between the corners is predominantly mediated
by the bulk modes, not by the edge states.
This is demonstrated by the $120^{\circ} - 120^{\circ}$ plots (the OA-OA setups) in figure~\ref{fig:trap_zz}
which shows the almost uniform high conductances across the whole interval of gate voltages.
In this way our data in terms of the conductance reflect the fact that the ZZ edge state does not exist
at and close to $120^\circ$ corners~\cite{Cuong_2013}.
The transport between the $120^{\circ}$ corners of the ZZ-edge structures is only marginally affected by the ZZ edge
states.

At the end of this section we conclude that
outside the narrow range of the ZZ-edge conductance window
the linear conductance of ZZ-terminated trapezoid is
typically vanishing unless the two electrodes are attached at the OA corners.
Outside the ZZ-edge conductance window
the contrast between the conductances of the OA-OA and AA-AA setups is several orders of magnitude.
The other similarity to the case of ac-edge terminated trapezoids is that
the results for the two AA-OA setups are very close each other (red and green plots in figure~\ref{fig:trap_zz}).
Contrary to the ac edges, the proximity of the two AA-OA curves is now found also for long-narrow structures;
additional examples of this feature can be found in SI on the case of rhomboidal flakes.
Although we do not analyse this effect we can say that it is related to the electronic modes which are used for the
transport: the edge states and the bulk states.
Finally, we again refer to appendix section~\ref{sec:1st3rdNN}, in particular figures~\ref{fig:1st_3rd_NN_zz_overall}
and~\ref{fig:1st_3rd_NN_zz_detail}, which use the linear scale on their conductance axes.
%
\subsection{Corners formed by different pair of edges: isosceles triangles}
%
Besides the trapezoids, we have addressed also graphene flakes of several other shapes, including
rhomboids, rhombi and equilateral triangles.
%
\begin{figure}[!t]
\centerline{\includegraphics[width=0.60\textwidth]{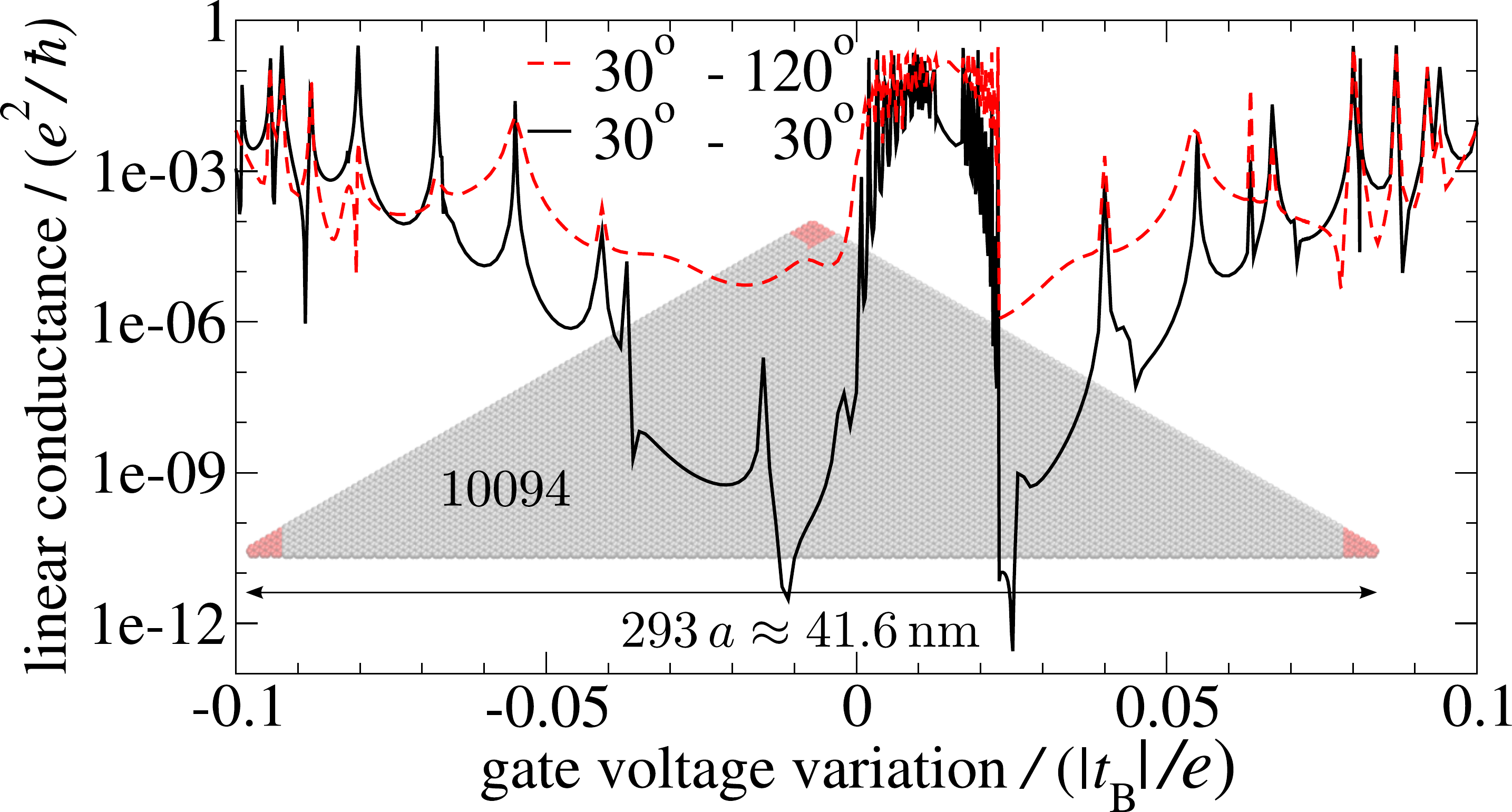}}
\caption{The linear conductance for a junction formed by an isosceles-triangle shaped graphene flake
with the electrodes attached at its corners.
The inset shows an image of the flake.
The size of the structure is $10\,094$ atoms.
The base edge is of the ac type while the edges along the sides are of the ZZ pattern.
The electrodes' thickness is $33$ monoatomic wires.
Other description is analogous to that in previous two figures.}
\label{fig:isoscel_tria_ac}
\end{figure}
%
All of these feature $60^\circ$ and $120^\circ$ angles at their corners.
Not surprisingly, they provide basically the same picture as the results for trapezoids described above.
Some of these results are show in SI.
As a different example here we consider a flake of an isosceles triangle shape with $30^\circ$ acute angles,
the ac edge at the base and the ZZ edges along its sides.
In other words, in the $30^\circ$ angle corners the two crossing edges are of the different types (ac and ZZ).
Based on the above findings, we intuitively expect a large contrast between the conductances
of the setups $30^\circ - 30^\circ$ and $30^\circ - 120^\circ$.
Quantitative results shown in figure~\ref{fig:isoscel_tria_ac} confirm this expectation,
with the exception of the isolated resonances and, more significantly, with the exception of
the $0.025\,|\tB|$ narrow high-conductance channel which is the signature of the
presence of the ZZ edge, although now only single one at each acute-angle corner.
We can expect that results for the case of a ZZ base would be similar to those in figure~\ref{fig:isoscel_tria_ac}.
%
\subsection{Contact size effect}
%
Naturally, the conductance depends on the size of the contact area which in our model is represented
by the electrode's thickness and is quantified by the number of the monoatomic wires per electrode.
%
\begin{figure}[!t]
\centerline{\includegraphics[width=0.60\textwidth]{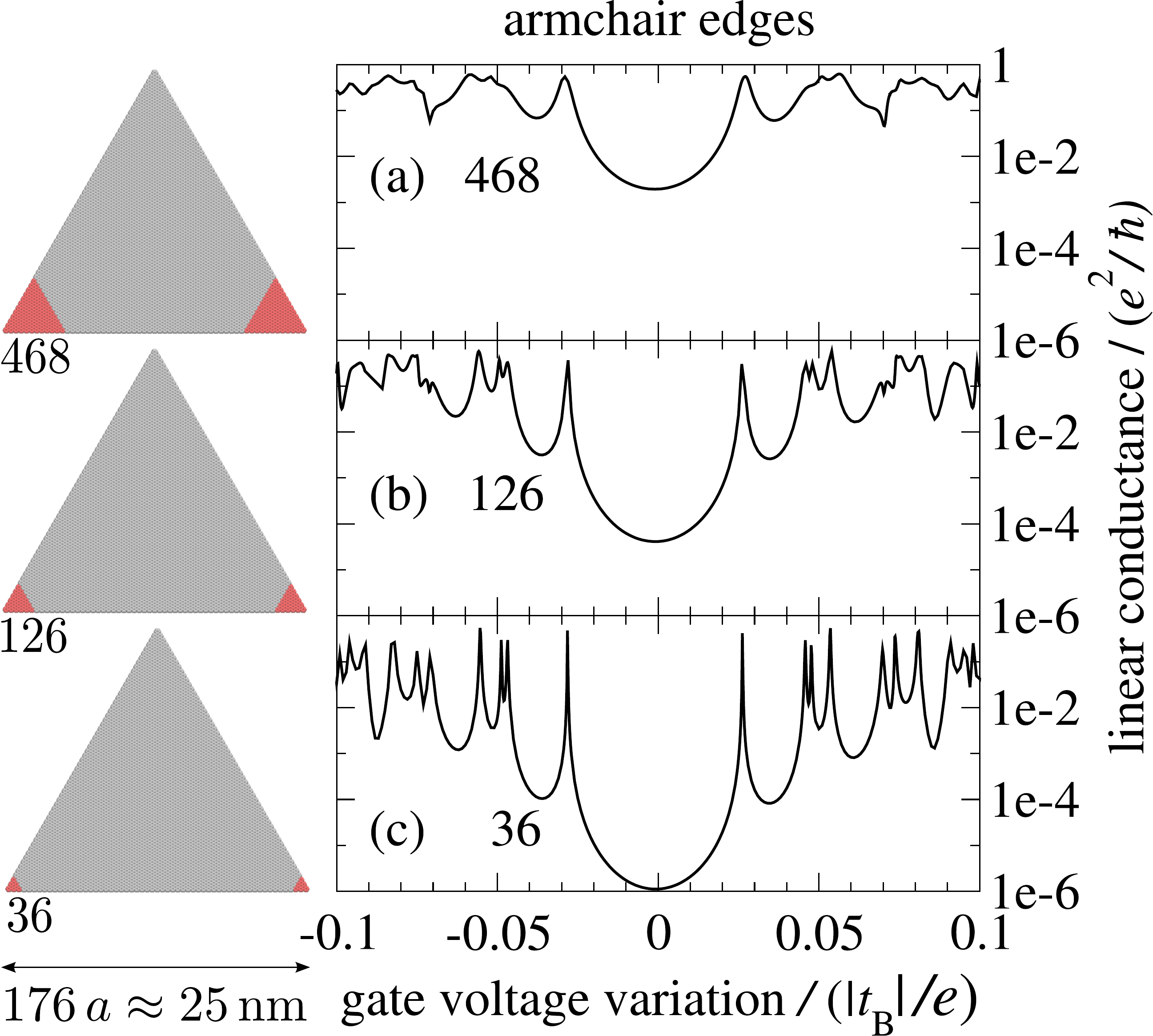}}
\caption{The linear conductances for different sizes of the contact areas (marked by red color).
The equilateral triangle-shaped flakes have ac edges.
Each triangle is of the same size, consisting of $10\,620$ atoms.
The numbers 36, 126 and 468 (both in graphs and at the triangles) denote the number of contacted atoms per corner,
i.e. the electrode's thickness in terms of the number of the composing wires.}
\label{fig:etria_ac}
\end{figure}
%
We have seen that graphene flake junctions can be highly conductive especially for the OA-OA contacts
even if the contact area corresponds to just 35 or 36 atoms.
For these contact sizes the AA-AA setups have usually been found relatively insulating for most of the applied gate
voltages, the exception being ZZ-edge mediated transport window.
In this section we study if and how the predominantly insulating behaviour is modified due to increased contact areas.
We opt to study this effect on equilateral-triangle shaped flakes.
The results are shown in figures~\ref{fig:etria_ac} and~\ref{fig:etria_zz}.
%
\begin{figure}[!t]
\centerline{\includegraphics[width=0.60\textwidth]{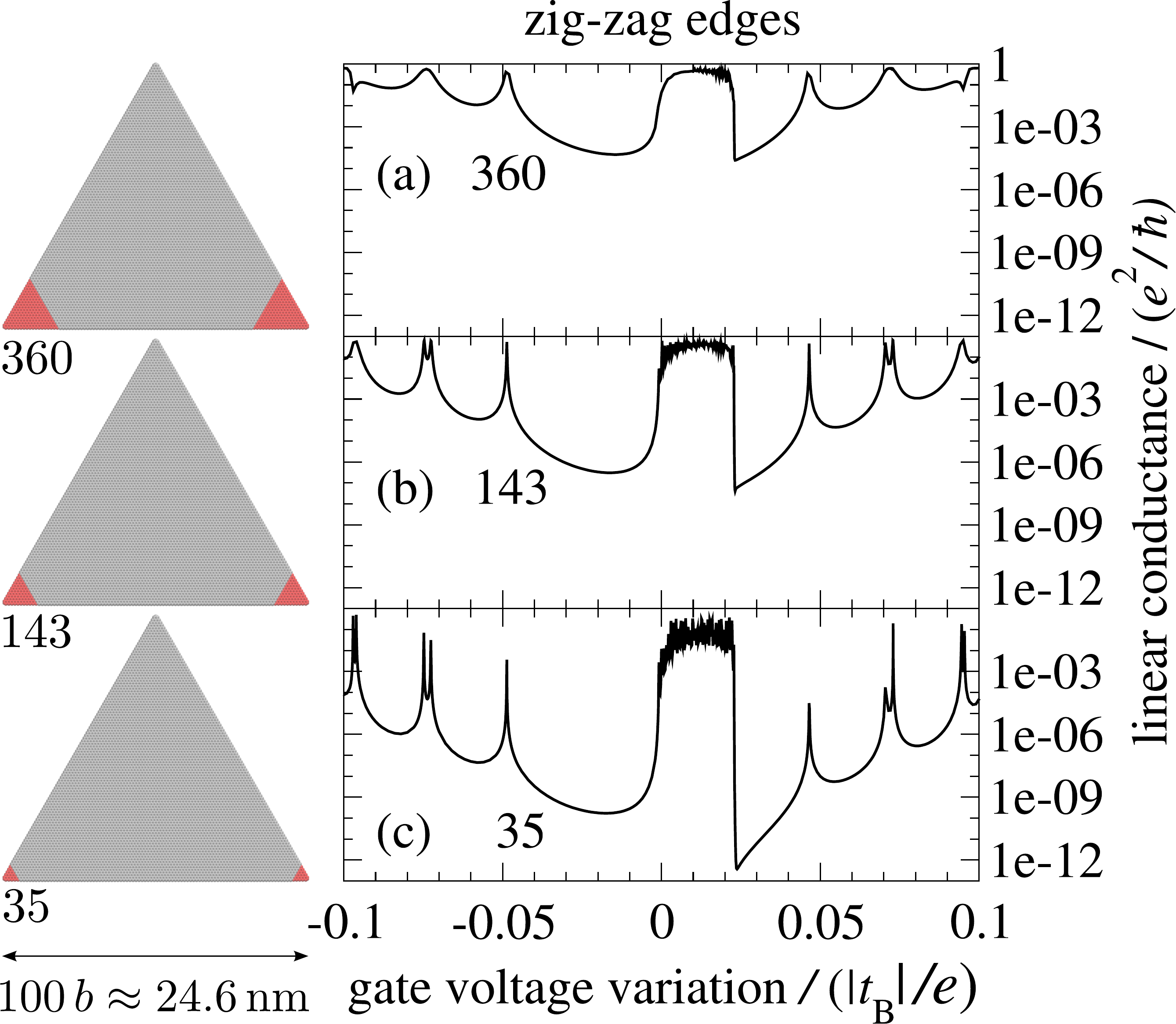}}
\caption{Same kind of results as in figure~\ref{fig:etria_ac} but now for the flakes with ZZ edges.
The triangle is composed of $10\,401$ atoms.
The numbers 35, 143 and 360 denote the number of contacted atoms per corner.}
\label{fig:etria_zz}
\end{figure}
%
For the ac-terminated structures we find that the increase of the contact areas leads to the broader peaks
of the conductance spectrum
while the peaks' maxima remain roughly unchanged.
The minima of the spectrum become systematically higher for thicker electrodes.
Although the conductance for the largest contact area ($468$ atoms) is significant at a whole range of elevated gate
voltages, the ratio of the conductance to the number of contacted atoms remains small compared to the cases
when contacts are made at the obtuse angles (figure~\ref{fig:trap_ac}, the blue curves).

In case of the ZZ-edge terminated triangles (figure~\ref{fig:etria_zz}) the conductance displays
the central ZZ-edge mode contribution and few additional very narrow resonances.
The spectrum within the high-conductance window does not change significantly with the thickness of the electrodes.
The remaining part of the calculated spectra in figure~\ref{fig:etria_zz}, while being rather asymmetric in comparison
to the ac case (figure~\ref{fig:etria_ac}), exhibits a similar dependence on the electrodes' thickness.
%
%
\section{\label{sec:LDOS} LDOS analysis}
%
%
The current-blocking \textit{vs.} conductive behaviour reported above can be understood by studying the electronic
modes of the graphene flakes.
We use the local density of states (LDOS) of isolated flakes as the tool.
%
\begin{figure}[!t]
\centerline{\includegraphics[width=0.45\textwidth]{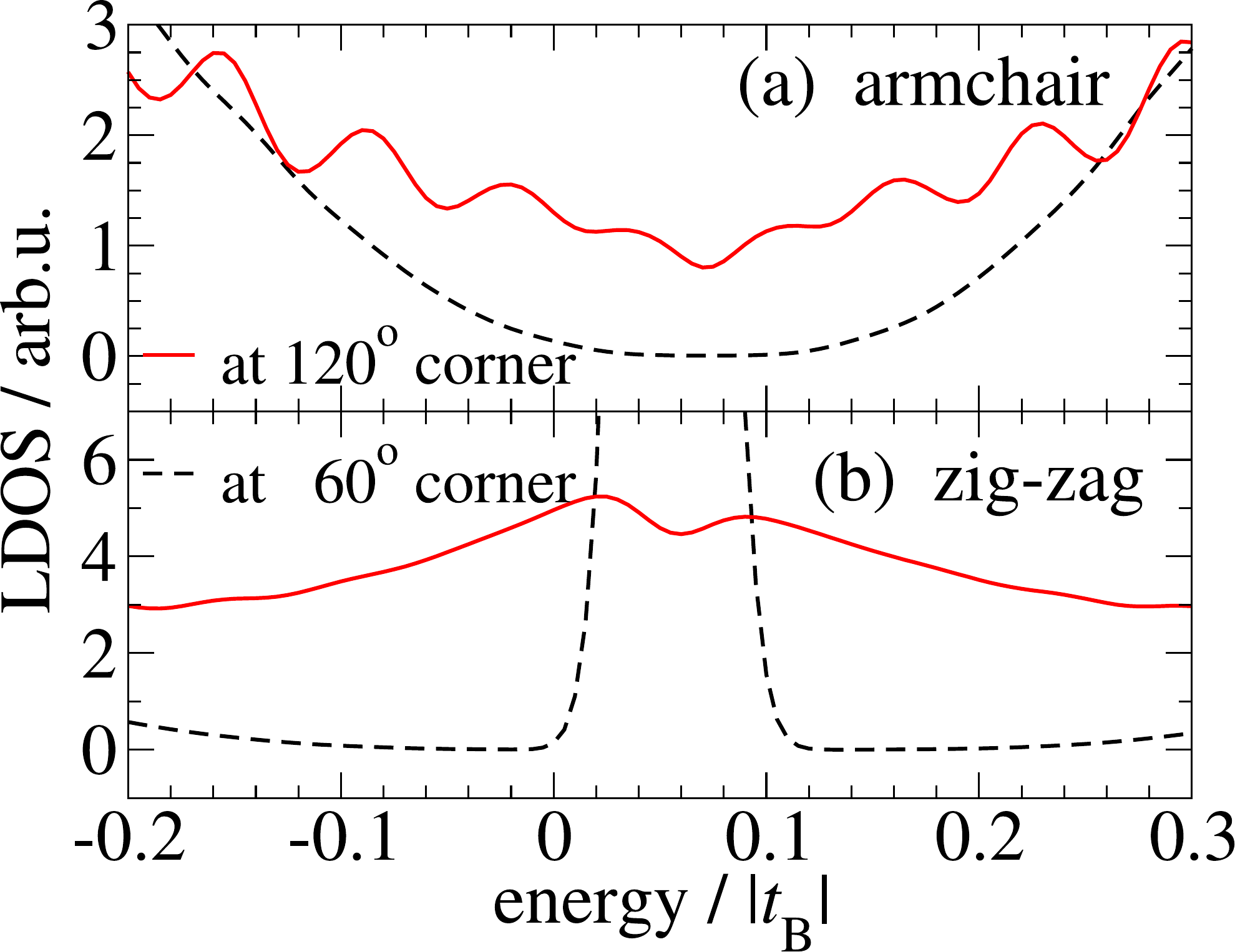}}
\caption{Local density of states (LDOS) in corners of isolated graphene structures.
The function is computed over the red-marked corner regions
of two trapezoidal graphene flakes.
(a)~LDOSs in the $60^\circ$ and $120^\circ$ corners of the armchair flake that is shown on
figure~\ref{fig:trap_ac},
the second structure from the top.
(b)~LDOSs in the corners of the zig-zag flake that is shown on
figure~\ref{fig:trap_zz},
the second structure from the top.
The legends referring to the angles apply globally within both graphs.
The LDOS curves are plotted as they come from calculations, without any additional shift like $\Delta\Vg$.}
\label{fig:LDOS}
\end{figure}
%
We start from the definition of the density of states of an isolated flake projected on an atomic site $l$:
\begin{equation}
\rho_l(E) = \sum_i |\langle l|\psi_i\rangle|^2 \delta(E - E_i)
\, ,
\label{eq:rho}
\end{equation}
with
$|l\rangle$ being the atomic orbital at the site,
$\psi_i$ is $i^\textrm{th}$ eigenfunction of the isolated flake and $E_i$ the associated eigenenergy.
We acquire LDOS information from a region of interest by summing up over corresponding sites:
\begin{equation}
\mathrm{LDOS}(E) = \sum_{l \in \mathrm{region}} \rho_l(E) \, .
\label{eq:LDOS}
\end{equation}
In the numerical evaluation the $\delta$-functions are replaced by normalised gaussians
\begin{equation}
\frac{1}{w\sqrt{2\pi}} \exp\left[-\left(\frac{E-E_i}{\sqrt{2}\,w}\right)^2\right]
\, ,
\end{equation}
with $w = 0.015\,|\tB|$.
The spectra are plotted in figure~\ref{fig:LDOS}.
Its graph~(a) shows LDOSs for the case of the ac-terminated trapezoid composed of 6324 atoms and displayed
in figure~\ref{fig:trap_ac}, the second structure from the top.
Graph~(b) of figure~\ref{fig:LDOS} shows LDOSs for the ZZ-terminated trapezoid composed of 6306 atoms and
displayed in figure~\ref{fig:trap_zz}, the second structure from the top.
The $l$-summation in equation~\eqref{eq:LDOS} runs over the red-marked atoms of the chosen corner
[36 atoms in case (a) and 35 atoms in case (b)].
In the ac case the LDOS at the $120^\circ$ corner (solid red line) is clearly larger than the LDOS at the
$60^\circ$ corner (dashed black line) within the central range of energies relevant for the electronic transport studied
here.
In the ZZ case [figure~\ref{fig:LDOS}(b)] the situation is more complicated because of the presence of the special
localised edge state~\cite{Nakada96} at zero energy and the corresponding peak in the LDOS.
Still, apart form the edge state contribution, the values of LDOS at the $120^\circ$ corner are again by far 
larger than the LDOS at the $60^\circ$ corner.
Contribution of the edge state to the electronic transport between corners of the flakes studied here is
specifically limited or suppressed in some cases:
this state does not exist at and close the $120^\circ$ angles of graphene flakes as it was shown
in~\cite{Cuong_2013} and as our analysis confirms, see discussion in section~\ref{ssec:ZZE}.
Our LDOS data for the ZZ case confirm the absence of the edge state:
the solid red plot in figure~\ref{fig:LDOS}(b) (corresponding to the $120^\circ$ angle)
reaches a maximum of only about $5$ on the given interval
while the dashed black curve (the $60^\circ$ angle) has its maximum at about $68$ (beyond the axis scale).

Finally we note that while in the graphs with $\Glin$ we have applied a uniform horizontal shift
$\Delta\Vg = +0.072\,|\tB|/e$ to calculated data (see section~\ref{sec:methods} for explanation),
no such a shift is used for the LDOS data.
%
%
\section{\label{sec:concl} Discussion and Conclusions}
%
%
In this work we studied stationary zero-temperature quantum transport through junctions formed by non-rectangular
graphene flakes with about nanometer or several nanometers thick electrodes attached at the corners of the flakes.
Such structures are perfectly compatible with hexagonal graphene lattice.
Typical shapes are rhomboids, rhombi, trapezoids and triangles.
They feature corners with angles being integral multiples of $30^\circ$.
Many of these structures have all their edges of the same kind: either armchair (ac) or zig-zag (ZZ).
Isosceles triangles with $30^\circ$ acute angles provide combinations of the ac and ZZ edges.
Studied flakes have sizes up to almost $20\,000$ carbon atoms.
Assuming a low bias voltage regime we computed the linear conductance as a function of a gate voltage applied
to the flake.
The computational methodology was based on a tight-binding model with hoppings up to the third nearest
neighbor included and on the scattering approach.
The later technique was employed in stationary calculations in our recent work~\cite{our_PRB_2014}.

Main finding of the present work is that the conductance strongly depends on the size of the angles at the
corners to which the pair of the electrodes is attached.
Especially for flakes with ac edges we can
say that junctions with both electrodes connected at obtuse-angle (OA) corners provide 1-5 orders of magnitude
higher conductance compared to setups with the electrodes attached at acute-angle (AA) corners.
The OA-OA junctions often provide high conductance values typical for metallic graphene nano-ribbons.
On the contrary, the pair of the electrodes attached at acute angles forms a relatively insulating junction.
The contrast between the conductances of OA-OA and AA-AA setups extends beyond one order of magnitude
within a $0.15\,|\tB|$ wide interval of the chemical potential tuned around the Fermi level,
$\tB$ being the nearest-neighbor hopping parameter of the tight-binding model.
Exceptions are a few narrow isolated resonances at which also AA-AA setups yield a high conductance.
Quantitative values depend on the dimensions of particular flake.

For graphene flakes with ZZ edges the situation is complicated by the existence of the well-know edge state
at the Fermi level.
This mode couples with the electrodes and provides the high conductance also for AA-AA setups,
but only within a $0.025\,|\tB|$ narrow interval of the chemical potential close to the Fermi level.
For the remaining part of the spectra in the ZZ cases we again find
a huge contrast between the conductances which can be even 1-2 orders of magnitude larger than for ac edges.
The spectral range in which the high contrast is found is also significantly wider that in cases of flakes
with the ac edges.

We have found that the reported behaviour does not change qualitatively within a range of electrodes' thicknesses,
from sub-nanometer values to several nanometers.
Another important feature of real samples is the presence of edge defects and their impact on the electronic transport.
It is known that especially AGNRs are more affected by the presence of the edge disorder.
On the contrary, flakes with ZZ edges are much more robust in this respect thanks to the presence
of the edge state~\cite{White07}, see also~\cite{Mucciolo09,Morpurgo11}.
Although we did not address this problem in a systematic manner,
we have at least examined two representative trapezoidal flakes with imperfect edges,
an ac-edge terminated trapezoid with random edge defects as well as a ZZ analogue.
We provide details in appendix~\ref{sec:defects} and additional material in SI.
Our findings fully confirm the aforesaid properties:
especially for graphene flakes with ZZ edges the effects reported in the present paper remain preserved
to a large extent even in the case of significant edge disorder.
In the case of ac edges we observe significant drop in the conductance of the junction which would otherwise be highly
conductive with perfect edges.
Still, even in the ac case, the conductance of the OA-OA setup remains to be several orders of magnitude larger
compared to the AA-AA setup within a range of gate voltages.
We expect that experimental confirmation of the effects reported here could be realised using proper carbon
nano-ribbon or nanotube electrodes, possibly providing also a mechanical support for the graphene flake.
%
%
\section*{Acknowledgements}
%
%
This work was supported in parts by
the Slovak Research and Development Agency under the contract No.~APVV-0108-11
and by
the Slovak Grant Agency for Science (VEGA) through grant No.~1/0372/13.
The structures were visualised using VMD~\cite{VMD}.
The author thanks Peter Bokes and Peter Marko\v{s} for stimulating discussions.
\appendix
%
%
\section{\label{sec:1st3rdNN} $1^\textrm{st}$ \textit{vs.} $3^\textrm{rd}$ nearest neighbor model}
%
%
All results in main text were obtained within the 3NNTBA.
%
\begin{figure}[!t]
\centerline{\includegraphics[width=0.58\textwidth]{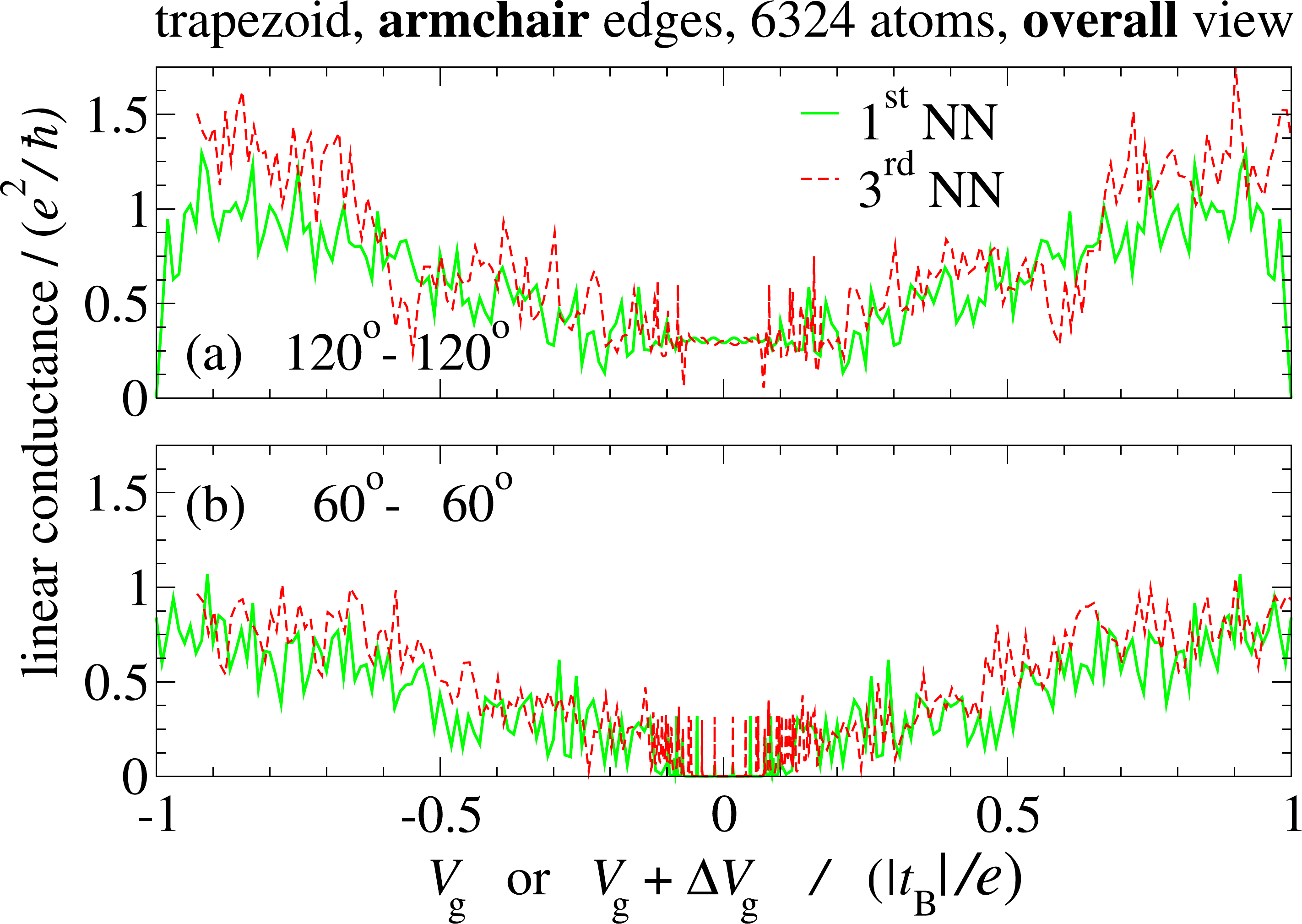}}
\caption{Linear conductances compared for the 1NNTBA (solid green line) and 3NNTBA (dashed red line) models.
The examined structure is the ac-terminated trapezoid composed of $6324$ atoms (figure~\ref{fig:trap_ac}).
The functions for the $3^\textrm{rd}$ NN model have been horizontally shifted
by $\Delta\Vg = +0.072\,|\tB|/e$ as specified in main text, section~\ref{sec:methods}.
Graph~(a) shows results for the electrodes attached at the $120^\circ$ angles.
Graph~(b) shows analogous results for the $60^\circ$ angles.
The models of the electrodes are identical to those used for results in figure~\ref{fig:trap_ac}.
See also figure~\ref{fig:1st_3rd_NN_ac_detail} for a detailed view on the low-$\Vg$ interval of the present graph.
Note that for figures within this section we use the linear scales also on vertical axes.}
\label{fig:1st_3rd_NN_ac_overall}
\end{figure}
%
%
\begin{figure}[!b]
\centerline{\includegraphics[width=0.58\textwidth]{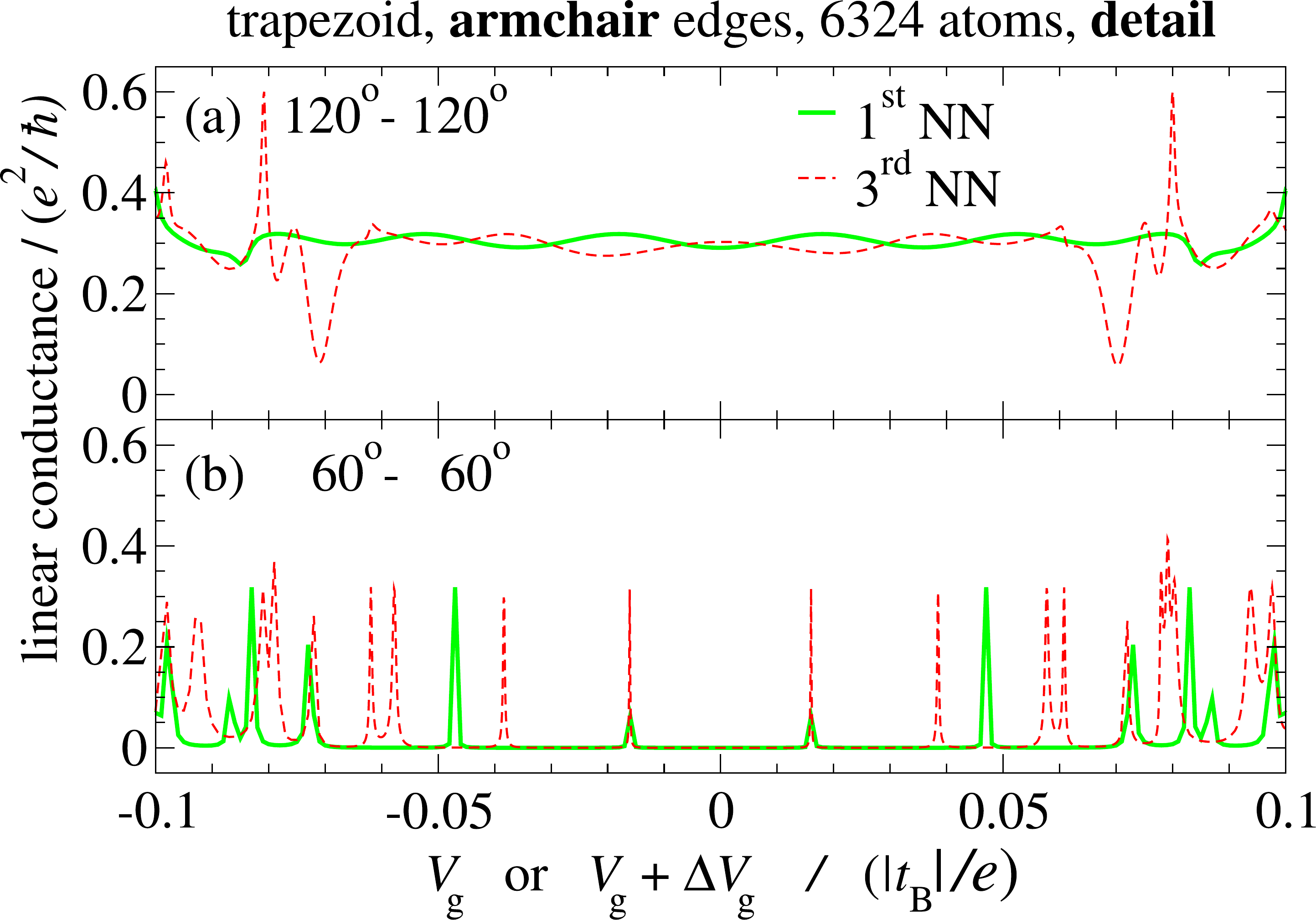}}
\caption{A detailed view of the results from figure~\ref{fig:1st_3rd_NN_ac_overall}.}
\label{fig:1st_3rd_NN_ac_detail}
\end{figure}
%
In addition we have performed several calculations in which TB hoppings up to only the first NN were included (1NNTBA).
In this section we compare results by 1NNTBA and 3NNTBA approaches and in this way
evaluate the effect of the extended hopping range on the conductance.
%
\begin{figure}[!t]
\centerline{\includegraphics[width=0.58\textwidth]{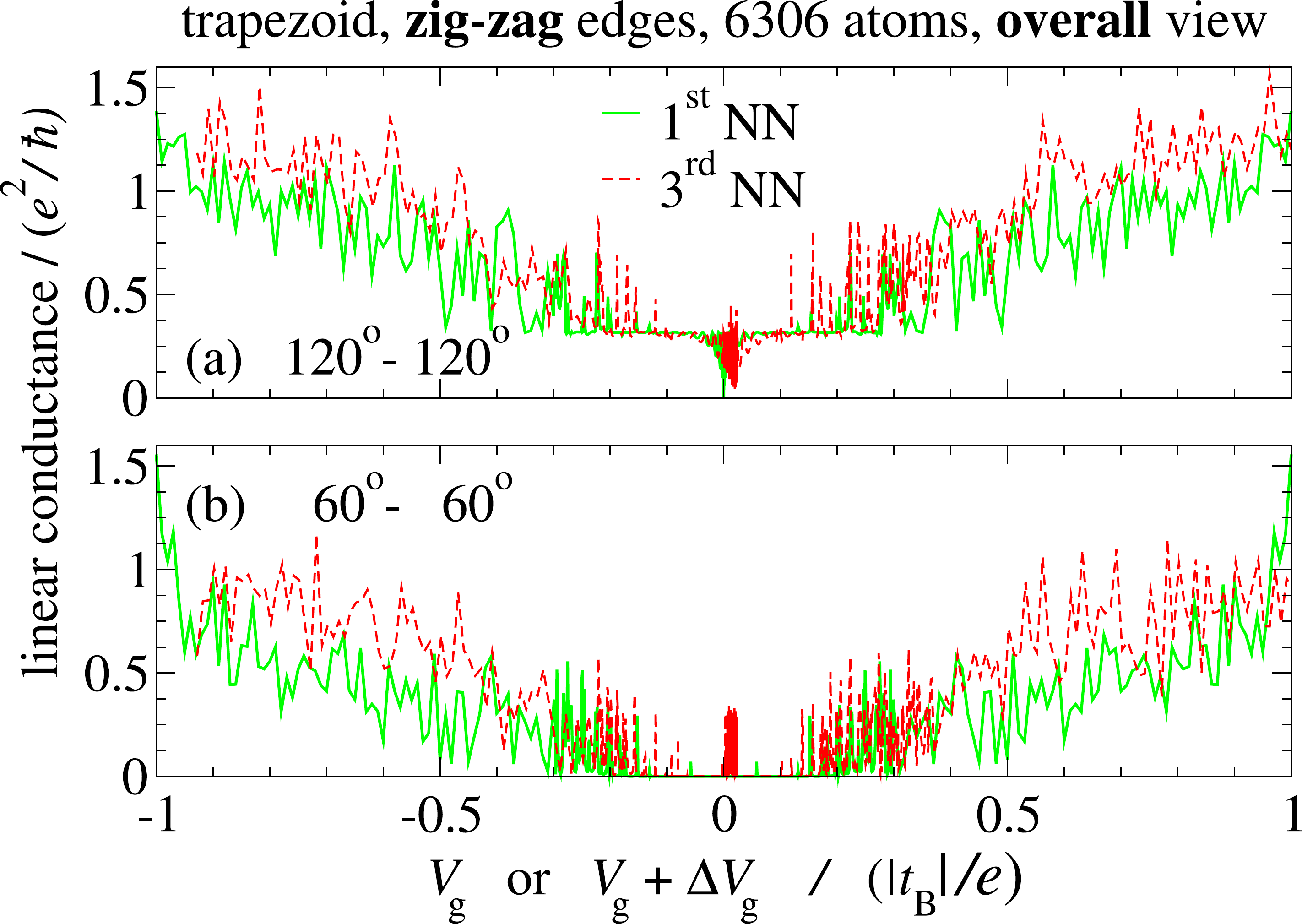}}
\caption{Linear conductances compared for the 1NNTBA (solid green line) and 3NNTBA (dashed red line) models.
The examined structure is the ZZ-terminated trapezoid composed of $6306$ atoms (figure~\ref{fig:trap_zz}).
The curves for the 3NNTBA model have been horizontally shifted
by $\Delta\Vg = +0.072\,|\tB|/e$ as specified in main text, section~\ref{sec:methods}.
Graph~(a) show results for the electrodes attached at the $120^\circ$ angles.
Graph~(b) shows analogous results for the $60^\circ$ angles.
The models of the electrodes are identical to those used for results in figure~\ref{fig:trap_zz}.
See also figure~\ref{fig:1st_3rd_NN_zz_detail} for a detailed view on the low-$\Vg$ interval of the present graph.}
\label{fig:1st_3rd_NN_zz_overall}
\end{figure}
%
%
\begin{figure}[!b]
\centerline{\includegraphics[width=0.58\textwidth]{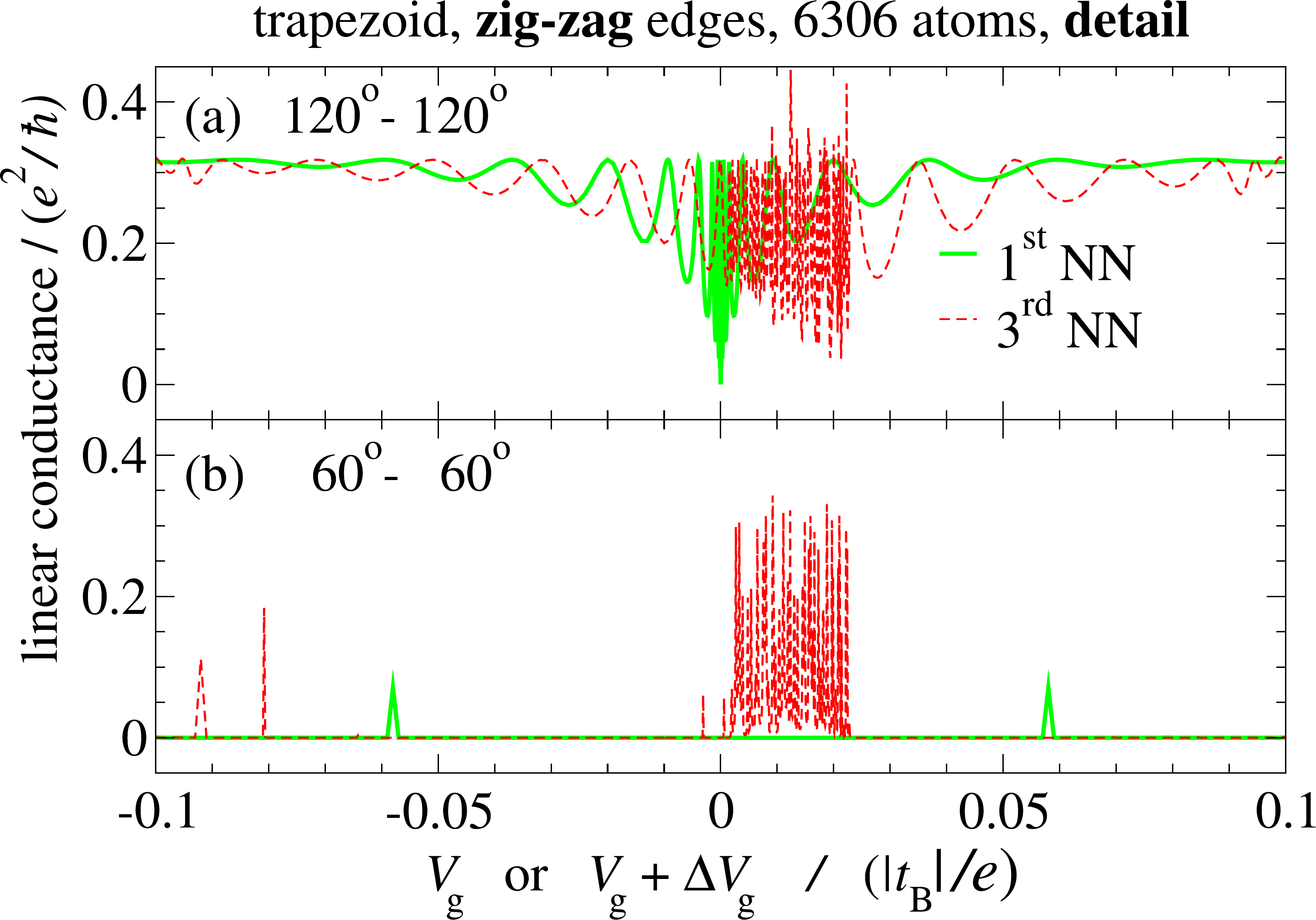}}
\caption{A detailed view of the results from figure~\ref{fig:1st_3rd_NN_zz_overall}.}
\label{fig:1st_3rd_NN_zz_detail}
\end{figure}
%
The wires forming the electrodes are always modelled within the 1NNTBA.
Examined systems are the ac-terminated trapezoid composed of $6324$ atoms (figure~\ref{fig:trap_ac})
and the ZZ-terminated one composed of $6306$ atoms (figure~\ref{fig:trap_zz}).
The comparisons are provided only for the AA-AA and OA-OA attachments.
As opposed to main text, in this appendix we use the linear scales on the conductance axes.

Overall view of the results for the ac case is shown in figure~\ref{fig:1st_3rd_NN_ac_overall}
where we display both the conductance within the 1NNTBA (the solid green plots)
as well as the results with the extended hoppings included (the dashed red plots).
The conductance spectrum in
the immediate neighborhood of the Fermi level is showed in a zoomed view in figure~\ref{fig:1st_3rd_NN_ac_detail}.
We see that the extended TB hopping range modifies the results only quantitatively.

We now turn to the case of the ZZ-edge terminated trapezoidal flake.
Results are shown in figures~\ref{fig:1st_3rd_NN_zz_overall} and \ref{fig:1st_3rd_NN_zz_detail}.
As we can see, in this case the extended TB hopping range has a noticeable effect on the conductances within
an about $0.02\,|\tB|/e$ narrow gate voltage window.
Most importantly, the AA-AA setup exhibits rapid oscillations of the conductance within this range
[figure~\ref{fig:1st_3rd_NN_zz_detail}(b), red plot],
in this way opening the high-conductance window.
Despite of this effect, in most of the $0.3\,|\tB|/e$ wide central gate voltage window the AA-AA setup remains
insulating [figure~\ref{fig:1st_3rd_NN_zz_overall}(b), red plot].
Similar rapid oscillations of the conductance are found also for the highly conductive OA-OA setup
[figure~\ref{fig:1st_3rd_NN_zz_detail}(a), red plot].
In this case the oscillations partially decrease the conductance within the narrow central window.
Still, the conductance remains in average significant as it was found also within the 1NNTBA.
%
%
\section{\label{sec:electrodes} Electrodes composed of interacting wires}
%
%
As specified in section~\ref{sec:methods}, each of the two electrodes in our model is composed of a bunch of identical
mutually non-interacting monoatomic wires;
see figure~\ref{fig:indep_wires} which shows a schematic graphical representation of such an electrode
and its contact to the graphene flake.
It may be questioned whether such a model of the electrodes is sufficiently realistic.
%
\begin{figure}[!t]
\centerline{\includegraphics[width=0.40\textwidth]{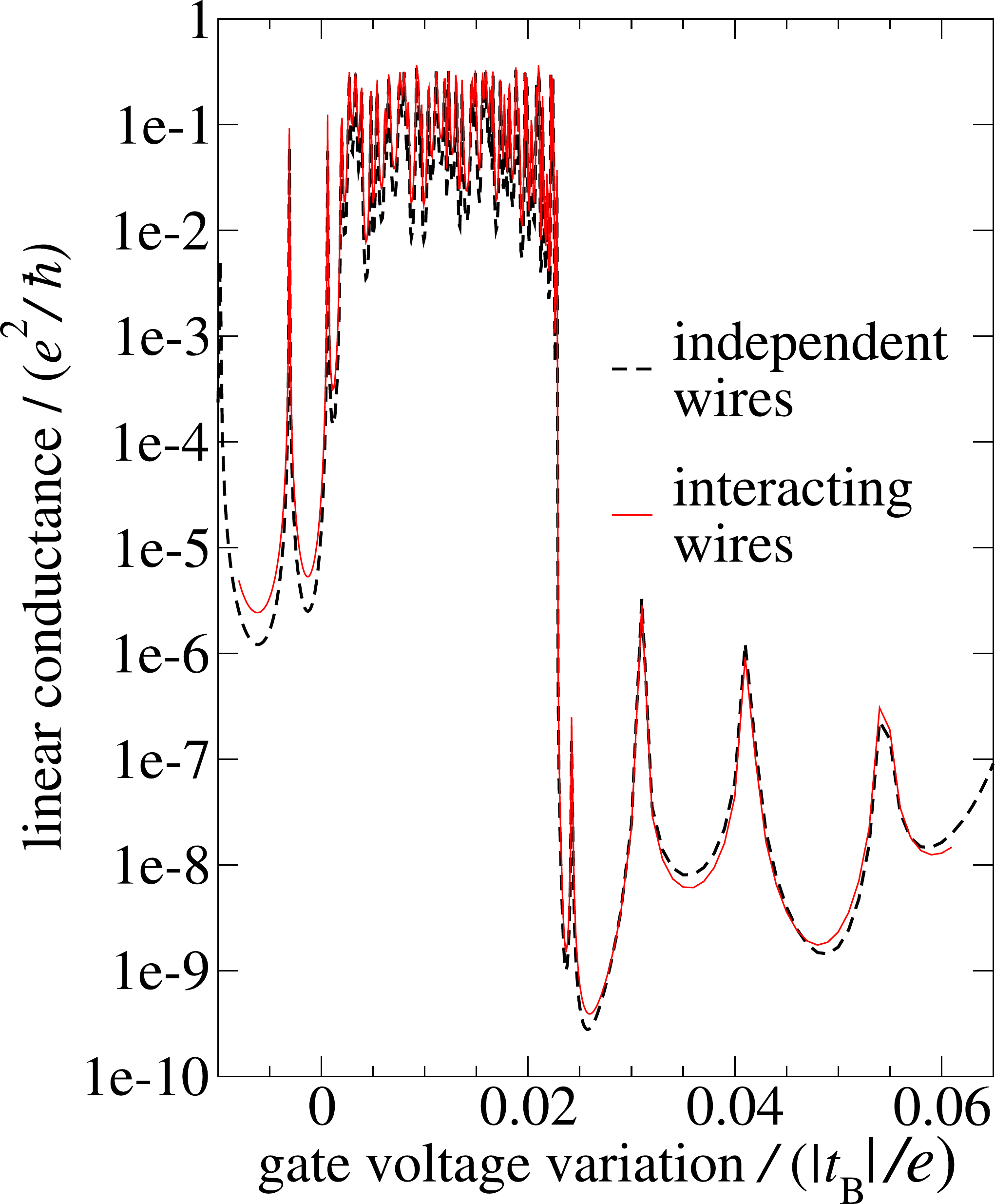}}
\caption{Linear conductance of the 6306-atoms large trapezoidal flake depicted on the left panel
of figure~\ref{fig:trap_zz}.
The electrodes are attached at the $60^\circ$ corners.
Dashed black plot in the present figure is just a detailed view of the black plot in figure~\ref{fig:trap_zz}(b).
Solid red plot in the present figure describes the same flake contacted to the different model of the electrodes.
Most importantly, the wires forming each of the two the electrodes are mutually coupled up to certain distance
from the flake.
See text in appendix~\ref{sec:electrodes} for more description.}
\label{fig:indep_vs_coupled_wires}
\end{figure}
%
We note that in theoretical studies of electronic transport in GNRs very simple models of electrodes
are often used, for example an electrode being just a continuation of the
GNR~\cite{Katsnelson06,Beenakker06,Perfetto10,Rosenstein09}.
Despite of the substantial difference between our model and the referenced ones we have shown~\cite{our_PRB_2014},
using AGNRs as examples, that our results are in semi-quantitative agreement with those employing the different models.
We provide similar evidence also in SI, see section~1 and figure~1 therein.

In addition here we perform an explicit test how would our result change if the wires were mutually interacting.
To achieve this we assume a model in which atoms of the wires within the $20\,a$ vertical distance from the flake
surface become mutually coupled using the 3NNTBA model with the same parametrisation as used for the graphene flake.
I.e. we assume an augmented central system formed by the flake and by the finite pieces of the electrodes (FPE).
The FPE are just those parts of the electrodes in which the wires are mutually coupled.
All couplings within the augmented system are treated on equal footing using the 3NNTBA model.
In this more complex model we still employ the semi-infinite mutually decoupled wires.
They are fixed to the ends of the FPE, not to the flake.
The couplings within the FPE, although not describing any real system, provide us with
a verification example whether or not the internal structure of the electrodes is important
for the effects studied in the present paper.
As the test case we choose the ZZ-edge terminated trapezoidal flake shown on the left panel of figure~\ref{fig:trap_zz},
the second structure from the top.
Results shown in figure~\ref{fig:indep_vs_coupled_wires} clearly demonstrate that the different model of the
electrodes has only a marginal impact on the results reported in our present work.
We can make the conclusion that the effects reported in the present paper
can be found for various types of electrodes assuming that they have order of nanometer thickness.
%
%
\section{\label{sec:defects} Impact of the edge disorder}
%
%
In order to qualitatively assess the impact of the irregular edges on the main effects reported in the present paper, we
perform calculations for two representative junctions.
%
\begin{figure}[!t]
\centerline{\includegraphics[width=0.49\textwidth]{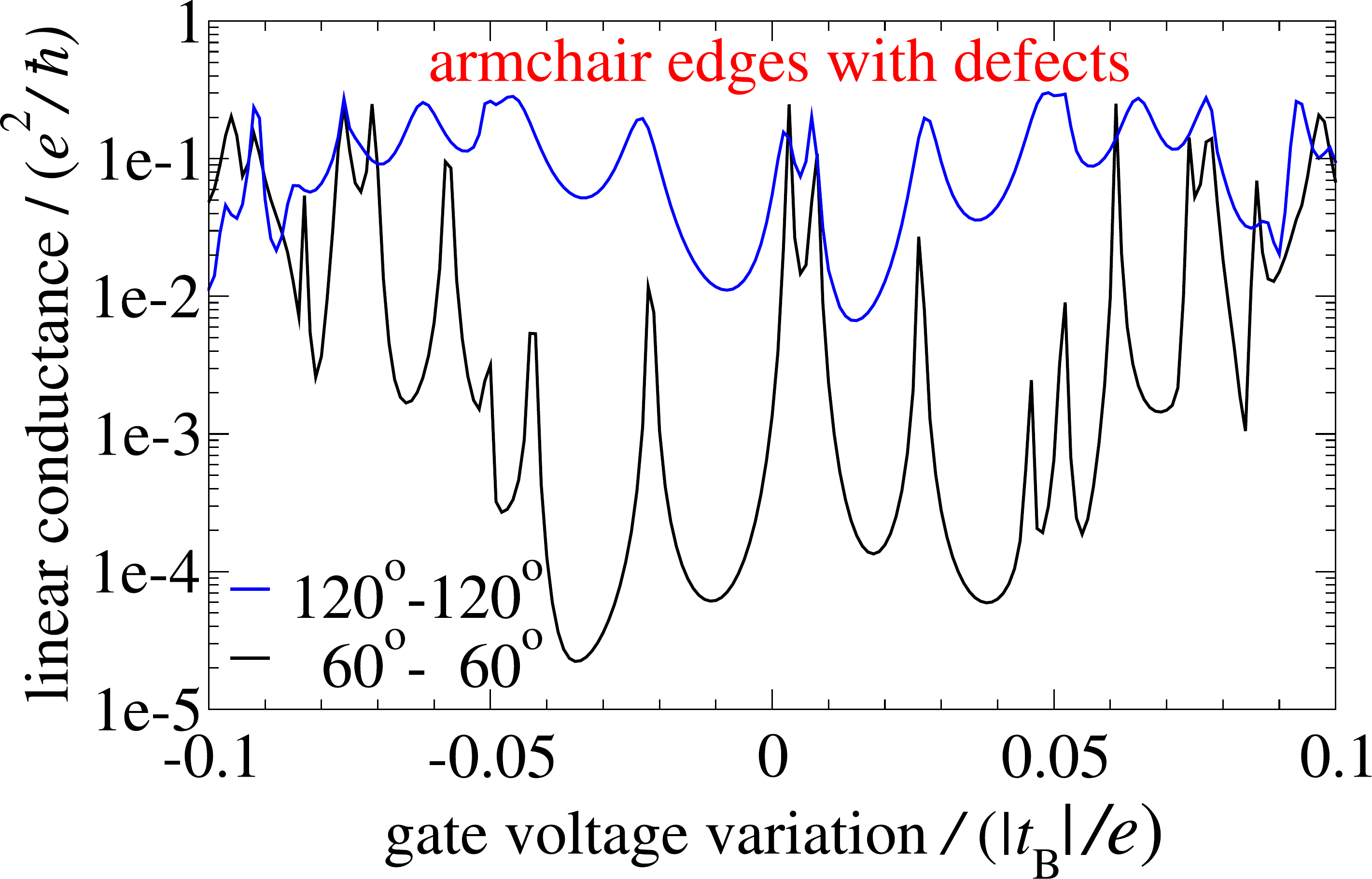}}
\caption{Linear conductance of the sample with the edge defects.
In the absence of the defects the flake would be identical to the one shown in the left panel
of figure~\ref{fig:trap_ac}, the second structure from the top.
The electrodes are 34 monoatomic wires thick.
Results for the AA-AA and OA-OA setups are shown only.
Graphs with direct comparisons between the perfect and imperfect border cases and an image of the structure
can be found in SI, figure~7 therein.}
\label{fig:irreg_edges_ac}
\end{figure}
%
First of them is based on the ac-edge terminated trapezoid shown in the left panel of figure~\ref{fig:trap_ac}, 
the second structure from the top.
After introduction of several random edge defect (including defects close to the corners) the structure consists
of 6245 atoms (an image is provided in SI, figure~7 therein).
The perturbations are made in such a way that every carbon atom at the border has either two or three nearest-neighbour
carbons.
Similarly as for the structures with regular edges, here we also assume that the edges are terminated by hydrogen
atoms~\cite{White07}.
All three bonding parameters $\tB$, $\tB'$ and $\tB''$ remain the same as for models with perfect edges.
Similarly we keep all equilibrium on-site energies at zeros.
Due to the defects present even at the corners the electrode's thickness has been slightly reduced
to 34 monoatomically thin wires for each of the two electrodes.
Calculated linear conductances for AA-AA and OA-OA setups are shown in figure~\ref{fig:irreg_edges_ac} and should
be compared to figure~\ref{fig:trap_ac}(b) which uses the same color coding.
Graphs directly comparing the perfect and imperfect-border cases can again be found in SI, figure~7 therein;
linear scales on vertical axes are used there.
%
\begin{figure}[!t]
\centerline{\includegraphics[width=0.49\textwidth]{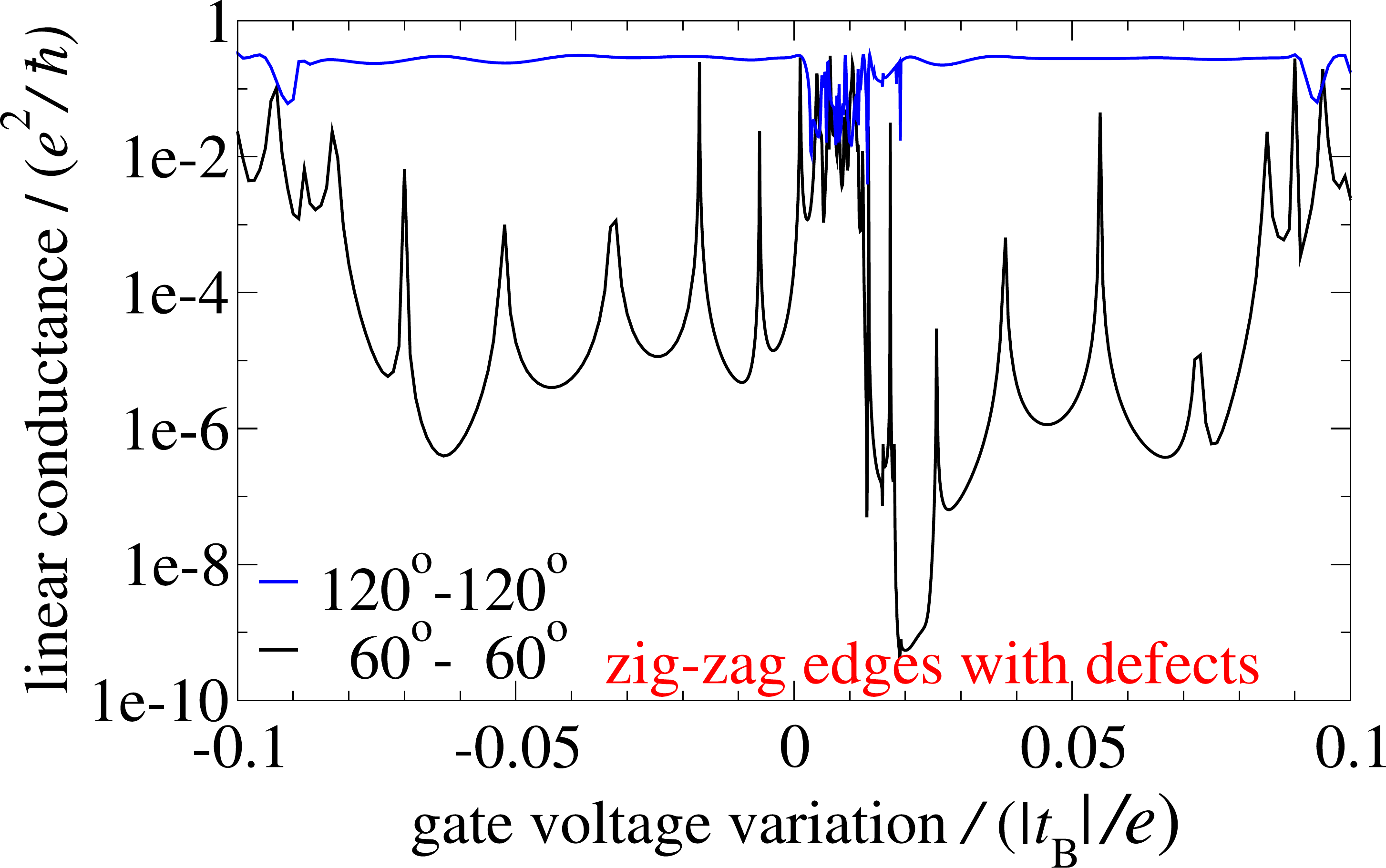}}
\caption{Linear conductance of the sample with the edge defects.
In the absence of the defects the flake would be identical to the one shown in the left panel
of figure~\ref{fig:trap_zz}, the second structure from the top.
The electrodes are 33 monoatomic wires thick.
Results for the AA-AA and OA-OA setups are shown only.
Graphs with direct comparisons between the perfect and imperfect border cases and an image of the structure
can  be found in SI, figure~8 therein.}
\label{fig:irreg_edges_zz}
\end{figure}
%
The impact of the edge disorder is multiple:
(i)~the relatively uniform $\Glin$ profile for the OA-OA setup now strongly varies with the gate voltage,
(ii)~the average magnitude of $\Glin$ has been decreased by about a factor of 10, i.e. the high conductance of the
OA-OA setup has been lost,
(iii)~the $\Glin$ profile for the AA-AA setup has got also a more oscillatory character,
(iv)~the $\Glin(\Vg)$ profile for the AA-AA setup now exhibits signs corresponding to ZZ edges;
see the central double peak in the black plot of figure~\ref{fig:irreg_edges_ac} just above the zero energy.
The latter feature is not surprising because some of the defect in the ac edges provide pieces of ZZ-type termination.
Despite of these modifications the conductance of the AA-AA setup is in average still several orders of magnitude below
those of the relatively conductive OA-OA setup.

The second junction with imperfect edges is based on the ZZ-edge terminated trapezoid shown in the left panel
of figure~\ref{fig:trap_zz}, the second structure from the top.
The flake with the defects is formed by 6219 atoms, the electrodes are 33 wires thick and the other conditions
and treatment are the same as in the ac case described above.
The result are presented in figure~\ref{fig:irreg_edges_zz} and should be compared to figure~\ref{fig:trap_zz}(b).
As in the ac case, additional graphs as well as an image of the flake can be found in SI, figure~8 therein.
We observe that the disorder introduced to ZZ edges has a noticeable impact on the linear conductance function.
This impact is smaller than in the ac case, fully in line with findings of~\cite{White07}.
Most significantly, the narrow window of the ZZ-edge state induced high conductance (just at the Fermi level)
becomes even narrower what is easily comprehended because the ZZ edges are now frequently interrupted by the defects.
Away from the Fermi level the conductance of the OA-OA setup is also high and, interestingly, almost unaffected
by the disorder [compare blue plots in figures~\ref{fig:trap_zz}(b) and \ref{fig:irreg_edges_zz}].
At these gate voltages the $\Glin(\Vg)$ profile of the (insulating) AA-AA setup is also little impacted
by the ZZ edge defects [compare black plots in figures~\ref{fig:trap_zz}(b) and \ref{fig:irreg_edges_zz}].

We can finally conclude that the main effects reported in the present paper are to a significant extent found
operational also for flakes with edge disorder.
\end{document}